\documentclass[aps,prm,reprint]{revtex4-2}

\usepackage{mathtools}
\usepackage{bm}
\usepackage{hyperref}
\hypersetup{colorlinks=true, citecolor=blue, urlcolor=blue, linkcolor=blue}

\begin{document}
\title{Monte Carlo Simulations of Crystal Defects in Open Ensembles}

\author{Flynn Walsh}
\email{walsh40@llnl.gov}
\author{Babak Sadigh}
\author{Joseph T. McKeown}
\author{Timofey Frolov}
\email{frolov2@llnl.gov}
\affiliation{Lawrence Livermore National Laboratory, Livermore, California 94550, USA}

\begin{abstract}

Zero- and two-dimensional crystal defects form in open statistical ensembles, such as the grand canonical, that are usually inaccessible with conventional simulation techniques.
This longstanding challenge is overcome with a new Hamiltonian Monte Carlo method that samples energy-biased gradual transformations.
The method enables free energy calculations for nonideal point defects and the direct prediction of finite-temperature interface structures.

\end{abstract}
\maketitle

Many physical systems can be modeled as open statistical ensembles containing variable numbers of particles ($N$) that are exchanged with reservoirs of constant chemical potential ($\mu$).
In crystals, open ensembles naturally describe point defects and grain boundaries (GBs), which form distinct phases for different values of $N$ with implications for macroscale properties \cite{tasker83,phillpot92,vonalfthan06,frolov13,frolov13a}.
The grand canonical ($\mu VT$) ensemble, which fixes volume ($V$) and temperature ($T$) in addition to $\mu$, receives the most attention, but solids usually experience fixed pressure ($P$) rather than volume.
This reality motivates the $\mu PT$ ensemble \cite{hill62,marzolino21,campa20,latella21}, which, though not appropriate for macroscopic systems, can represent atomic-scale solids within a region of convex free energy \cite{campa18}.

As in any other ensemble, $\mu PT$ and $\mu VT$ properties can be calculated by Monte Carlo (MC) sampling.
Standard Markov chain MC (MCMC) samples microstates by iteratively trialing mutations that are accepted with probability reproducing the desired energetic distribution \cite{hastings70}.
The simplest way to mutate $N$ is to randomly insert and delete particles, which, as a reversible procedure, can guarantee correctness by satisfying the condition of detailed balance.
In dense solids and liquids, however, naively adding or removing particles creates high-energy defects that are typically sampled with vanishing probability.
Some degree of structural relaxation is required to generate insertion and deletion trials that will be accepted at practical rates, but energy minimization is generally irreversible and therefore not well motivated.

Directly jumping among distinct structures may be possible in certain situations, but constructing these moves requires significant effort in practice \cite{tanguy24}.
As another approach, equilibrium molecular dynamics (MD) simulations can gradually add and remove particles by integrating atomic fractions as an additional degree of freedom with corresponding equations of motion \cite{cagin91,cagin91a,boinepalli03,eslami07}.
Similar techniques have been employed within MC frameworks \cite{shi07,rahbari21}, but the presence of unphysical partial particles is less than desirable in both cases.
In solids, this approach has been primarily considered for structure optimization rather than thermodynamics \cite{phillpot92}.

However, the promise of gradual insertions and deletions via MD can be fully realized using the flexibility of MCMC, which permits any moves that satisfy statistical balance.
For instance, in canonical ($NVT$) MC simulations, deterministic and reversible microcanonical ($NVE$) MD can generate trial states with probabilities corresponding to the choice of initial velocities.
This technique, which is known as Hamiltonian or hybrid MC \cite{duane87,mehlig92,betancourt17}, can also sample nonphysical trajectories.
For example, the addition of a fourth dimension has been used to move pair-interacting particles between a liquid and a reservoir \cite{belloni19,kim23}.
Related techniques have also been applied to biomolecular systems \cite{melling23}.

This work explores the addition and removal of solid atoms through the trajectories of a time-dependent Hamiltonian.
Atoms are not truly created or destroyed, but transformed between ``real" and ``fictitious" particles in a canonical representation of the grand canonical problem.
Specific transformations are attempted with a bias toward deleting high-energy atoms and inserting fictitious particles with low hypothetical energies.
Together, these techniques enable practical, general, and rigorous simulations of dense solids with variable numbers of atoms.
After validation and optimization, the method is used to calculate the free energies of arbitrary point defect structures with nonideal configurational entropies.
The remainder of the study demonstrates how $\mu PT$ MC can directly predict the finite-temperature structure of crystal GBs, providing a new approach to studying interfaces and their phase transitions.

In brief, $\mu VT$ MC was performed through the following procedure:
\begin{enumerate}
  \setcounter{enumi}{-1}
  \item Overlay $N$ real particles with $M-N$ noninteracting fictitious particles to create an $MVT$ ensemble.
  \item With fixed probabilities, choose to trial displacement ($W^0$), deletion ($W^{-}$), or insertion ($W^+$). \label{item1}
  \item \label{item2}
  \begin{enumerate}
    \item If deleting, select real atom $i$ with biased probability $W^{-i} \propto e^{\beta^- U_i}$ for some $\beta^-$.
    \item If inserting, select fictitious particle $j$ with biased probability $W^{+j} \propto e^{-\beta^+ U_j^{+}}$ for some $\beta^+$.
  \end{enumerate}
\item Assign random nonphysical momenta, $\bm{p}^M$, to all particles with probability density $W^{p} \propto e^{-\beta \bm{p^M} \cdot \bm{p^M} /2m}$
\item Update particle positions along a Hamiltonian trajectory, which gradually but fully transforms any particle selected in step \ref{item2}.
\item Calculate the probability of a reverse trial and accept or reject per the Metropolis-Hastings criterion. Return to step \ref{item1}. \label{item5}
\end{enumerate}
$\mu PT$ MC \cite{latella21} was performed similarly with the addition of volume-scaling trials in step \ref{item1}, which proceed directly to step \ref{item5} without otherwise updating positions.

Fictitious particles are introduced as a convenient way to identify sites where insertions are more likely to succeed.
They do not interact with real particles or each other, so the potential energy of a combined $M$-particle system, $U(\bm{x}^M)$, is determined by the positions of the $N$ real particles, $\bm{x}^{N \subset M}$.
However, the combined system contains many copies of each $N$-particle microstate, corresponding to all the ways to choose $N$ real particles from $M$ total, $M \choose N$ \cite{rowley75}.
Therefore, sampling the $M$-particle system with probability $\pi(\bm{x}^{N \subset M})$ approximates
\begin{equation}
  \label{eq:dxM}
  \int \pi(\bm{x}^{M}) \, d\bm{x}^M = \sum_{N=0}^M \frac{M!}{N!(M-N)!} \int \pi(\bm{x}^{N \subset M}) \, d\bm{x}^M.
\end{equation}
Since $\pi(\bm{x}^{N \subset M})$ depends only on real positions, the integrals over fictitious coordinates simplify as $\int d\bm{x}^{M-N} = V^{M-N}$.
It is now useful to introduce $\Lambda = \sqrt{\beta h^2 / 2 \pi m}$ for $\beta=1/k_B T$, mass $m$, and Planck's constant $h$. If
\begin{equation}
  \label{eq:pNM}
  \pi(\bm{x}^{N \subset M}) = \frac{(M-N)!}{M!} \frac{V^{N-M}}{\Lambda^{3N}} e^{-\beta\left[U\left(\bm{x}^{N \subset M}\right) - \mu N\right]},
\end{equation}
then 
\begin{equation}
  \label{eq:Z}
  \int \pi(\bm{x}^{M}) \, d\bm{x}^M = \sum_{N=0}^M \frac{1}{N!} \frac{1}{\Lambda^{3N}} \int e^{-\beta \left[U\left(\bm{x}^{N\subset M}\right) - \mu N\right]} \, d\bm{x}^N,
\end{equation}
which is the configurational partition function of the $\mu VT$ ensemble for $N<M$.
This means that sampling an $MVT$ system with $\pi(\bm{x}^{N \subset M})$ given by Eq. (\ref{eq:pNM}) models a $\mu VT$ ensemble with negligible $\pi(\bm{x}^{N > M})$.
The $\mu PT$ ensemble similarly motivates $MPT$ proxy systems (see Appendix \ref{app:muPT}).

Sampling was performed in the standard Metropolis-Hastings framework \cite{hastings70}.
MC trials evolved the system from $\bm{x}_a$ to $\bm{x}_b$ according to probability density $W_{ab}$.
Moves were accepted with probability $A_{ab}$ to achieve transition rates $K_{ab} = W_{ab} A_{ab}$ that ensured $\pi(\bm{x}_a) K_{ab} = \pi(\bm{x}_b) K_{ba}$, specifically,
\begin{equation}
  \label{eq:Aab}
  A_{ab} = \min\left(1,\frac{W_{ba}}{W_{ab}} \frac{ \pi(\bm{x}_b)}{\pi(\bm{x}_a)} \right).
\end{equation}
From Eq. (\ref{eq:pNM}), the microstate probability ratios in deletion and insertion trials were, respectively,
\begin{align}
  \medmuskip=0mu
  \label{eq:pdelete}
  \frac{\pi(\bm{x}_b^{N-1\subset M})}{\pi(\bm{x}_a^{N\subset M})} &= \frac{M-N+1}{V\Lambda^{-3}} e^{-\beta\left[ U\left(\bm{x}_b^{N-1\subset M}\right) - U\left(\bm{x}_a^{N\subset M}\right) + \mu\right]},\\
  \label{eq:pinsert}
  \frac{\pi(\bm{x}_b^{N+1\subset M})}{\pi(\bm{x}_a^{N\subset M})} &= \frac{V\Lambda^{-3}}{M-N} e^{-\beta\left[ U\left(\bm{x}_b^{N+1\subset M}\right) - U\left(\bm{x}_a^{N\subset M}\right) - \mu\right]}.
\end{align}

The other half of Eq. (\ref{eq:Aab}) depends on trial probability densities $W_{ab}$ and $W_{ba}$, which are composites of several steps.
First, the type of MC was chosen: displacements, deletions, and insertions were attempted with, respectively, probabilities $W^{0}$, $W^{-}$, and $W^{+}$.
Deletion and insertion trials also selected a particle for gradual transformation.
Most simply, random $W_a^{-i}=1/N$ and $W_a^{+j}=1/(M-N)$, but more efficient choices are examined later.
New positions were then generated from a Hamiltonian trajectory with probability density $W_{ab}^{x}$, as discussed below.
Altogether, the relative probability densities of displacement, deletion, and insertion trials were, respectively,
\begin{align}
  \label{eq:Wab0}
  W_{ab}^{0} &= W^{0} W_{ab}^{x} \\
  \label{eq:Wabdel}
  W_{ab}^{-i} &= W^{-} W_a^{-i} W_{ab}^{x} \\
  \label{eq:Wabins}
  W_{ab}^{+j} &= W^{+} W_a^{+j} W_{ab}^{x}.
\end{align}

Hamiltonian trajectories, which are integrated by Verlet MD, provide a flexible framework for sampling particle positions \cite{mehlig92,betancourt17}.
Displacements were generated using the standard time-independent Hamiltonian, $H_N = \bm{p}^M \cdot \bm{p}^M /2m + U(\bm{x}^{N \subset M})$.
In insertion and deletion trials, the number of real particles was incremented from $N$ to $N\pm1$ by evolving the system under the time-dependent Hamiltonian
\begin{equation}
H_{\pm}(t) = \left[1-\lambda(t)\right] H_N + \lambda(t) H_{N\pm1}
\end{equation}
where $\lambda(t) = (t-t_a) / (t_b - t_a)$ for a trajectory from $t_a$ to $t_b$.

For initial state $\bm{x}^M_a$, the final state $\bm{x}^M_b$ depends only on the choice of initial momenta, $\bm{p}_a^M$, which are distinct from the physical values represented by $\Lambda^{-3N}$ in the configurational partition function of Eq. (\ref{eq:Z}).
Initial momenta were chosen to resemble those at the end of trajectories, which approach Maxwell-Boltzmann distributions.
The cumulative momentum probability density, $W^p_a$, can be related to position trial density, $W^x_{ab}$, by considering discrete probabilities $W_{ab}^x d\bm{x}_b^M = W_a^p d\bm{p}_a^M$.
Altogether,
\begin{equation}
\label{eq:Wab}
  W_{ab}^x = \frac{d\bm{p}_a^M}{d\bm{x}_b^M} W_a^p = \frac{d\bm{p}_a^M}{d\bm{x}_b^M} e^{-\beta{\bm{p}_a^M \cdot \bm{p}_a^M / 2m}}.
\end{equation}

If, after a successful trial from $\bm{x}_a^M$ to $\bm{x}_b^M$, new momenta were drawn as $-{\bm{p}}_b^M$, then the trajectory would run exactly backward, recovering $\bm{x}_a^M$ (see Supplemental Material \cite{NOTE1}).
Therefore, the probability density of a reverse trial can be similarly determined as 
\begin{equation}
  W_{ba}^x = \frac{d\bm{p}_b^M}{d\bm{x}_a^M} e^{-\beta (-\bm{p}_b^M) \cdot (-\bm{p}_b^M) / 2m }.
\end{equation}
Since Hamiltonian trajectories conserve the volume element $d\bm{x}^M d\bm{p}^M$ \cite{jarzynski97}, the differentials cancel in Eq. (\ref{eq:Aab}).
The momenta of untransformed fictitious particles, which retain constant magnitude, can also be neglected.

If certain particles are significantly easier to insert or delete than others, then random transformations can be highly inefficient.
Instead, deletion trials were biased to select particles with high energies.
Atoms with initial energies $U_i(\bm{x}_a^{N \subset M})$ were specifically selected with probability 
\begin{equation}\label{eq:Wdel}
W_{a}^{-i} = \frac{ e^{\beta^{-}  U_i(\bm{x}_a^{N \subset M})} }{ \sum_{i}^{N} e^{\beta^{-} U_{i}(\bm{x}_a^{N\subset M})}},
\end{equation}
where $\beta^{-}$ is an adjustable parameter.
This distribution temperature controls the strength of the selection bias, which reverts to random in the limit of $\beta^{-} = 0$.

Insertions can be guided similarly.
Even though fictitious particles experience no interactions, their hypothetical energies upon direct insertion were calculated based on the positions of real atoms.
In an insertion trial, fictitious particle $j$ with hypothetical energy $U_j^{+}(\bm{x}_j,\bm{x}_a^{N \subset M})$  was selected with probability
\begin{equation}\label{eq:Wins}
  W_a^{+j} = \frac{ e^{-\beta^{+}  U_j^{+}(\bm{x}_j,\bm{x}_a^{N \subset M}) } }{ \sum_{j}^{M-N} e^{-\beta^{+} U_{j}^{+}(\bm{x}_j,\bm{x}_a^{N \subset M}) }},
\end{equation}
using a distinct adjustable bias temperature, $\beta^{+}$.

The method was implemented in the LAMMPS code \cite{thompson22} in a manner that will be detailed in a future publication.
The procedure's correctness was validated by a series of tests in elemental Cu \cite{NOTE1}.
All simulations were initialized with an equal number of real and fictitious particles, i.e., $M=2N_0$.
This choice proved consistently effective but should be more carefully analyzed in future studies.
While liquids could be simulated with a wide range of bias temperatures, solids were most efficiently sampled with $\beta^- = \beta$ and $\beta^{+} \approx \beta/5$.
Using these parameters, $\sim100$ step trajectories had acceptance rates of at least a few percent, which is sufficient for practical simulations.
Though this study considered simple interatomic potentials based on the embedded-atom method, almost all calculations could have been performed with much more expensive models.

\begin{figure}
  \includegraphics[width=\linewidth]{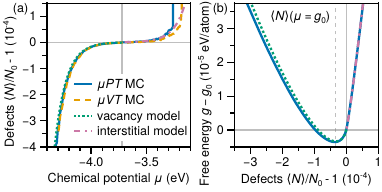}
\caption{
  \label{fig:solid}
  $\mu VT$ and $\mu PT$ MC simulations of fcc Cu at 1200\,K. 
  (a) Interstitial or vacancy concentrations, represented in terms of excess atoms per site, $\langle N \rangle / N_0 - 1$, as a function of chemical potential.
  $\mu PT$ calculations were also fit to an analytical model of free energy (see text).
  (b) Gibbs free energy per atom from integrating $dG = \mu dN$ with reference to the free energy of the perfect crystal at $N_0$.
  The dashed line indicates that $\langle N \rangle(\mu=g_0)$ effectively coincides with minimum $g$, validating the approximation of $\mu \simeq g_0$ for calculating the equilibrium vacancy concentration of an $NPT$ system.
}
\end{figure}

As an initial demonstration, fcc Cu was simulated at 1200\,K to study point defects.
Figure \ref{fig:solid}(a) plots the equilibrium number of atoms, which is measured in terms of interstitial or vacancy defects, across a range of chemical potentials in both $\mu VT$ and $\mu PT$ calculations.
Defect concentration is measured as $N/N_0 - 1$, where $N_0$ is the number of sites.
While the crystal remained stable with a large number of vacancies, interstitials led to the formation of a new atomic plane corresponding to the sharp jump in $\langle N \rangle$, which occurred at a lower interstitial concentration in $\mu PT$ simulations.

These data are interesting because they can provide the complete free energy of a crystal with point defects.
The Helmholtz ($F$), and hence Gibbs ($G=F+PV$), free energies of solids can be calculated by integration from a reference Einstein crystal \cite{frenkel84,freitas16}, but this approach poorly describes defects at even moderate temperatures due to diffusion.
Free energies can be more accurately integrated from low temperatures \cite{foiles94} or modeled with ideal configurational entropy \cite{zhang18}, but both approaches require the assumption of a specific defect configuration, e.g., a single monovacancy.
While arbitrary defect structures can form in large-scale calculations with free surfaces \cite{mendelev09}, calculating their free energies is not straightforward.
For $\mu PT$ simulations of a given lattice, the observation that $\mu = ( \partial G / \partial N)_{T,P}$ motivates an alternative approach of integrating $G(N)=G(N_0) +\int_{N_0}^{N} \mu dN'$ within the region of stability.

In reality, a crystal exposed to vacuum will minimize its Gibbs free energy by forming internal defects, which grow or shrink the lattice while conserving the total number of atoms.
For a $\mu PT$ simulation representing part of an extended $NPT$ crystal, this condition is equivalent to minimizing $g=G/N$.
Using numerically integrated $G$, Fig. \ref{fig:solid}(b) plots $g$ against $\langle N \rangle/N_0 - 1$ to reveal a minimum corresponding to the equilibrium vacancy concentration of an extended $NPT$ crystal.
In contrast to fluids, this is the only point where $\mu = g$ in a solid with point defects \cite{mullins85}. 
However, Fig. \ref{fig:solid}(b) indicates that the absolute difference between $g_{\rm{min}}$ and $g_0$ is small enough that the extended $NPT$ crystal can be approximated using $\mu = g_0 \simeq g_{\rm{min}}$.
As validation, the vertical line in Fig. \ref{fig:solid}(b) indicates that $\langle N \rangle (\mu=g_0)$ coincides with $g_{\rm{min}}$ within the resolution of the simulations.
Defect concentrations were also fit to a model of ideal configurational entropy (see Appendix \ref{app:ideal}), which largely reproduces the $\mu PT$ calculations in Fig. \ref{fig:solid}.
This is not surprising given the small numbers of vacancies and interstitials, but nonetheless validates $\mu PT$ MC as a viable tool for studying crystal defects.

\begin{figure}
  \includegraphics[width=\linewidth]{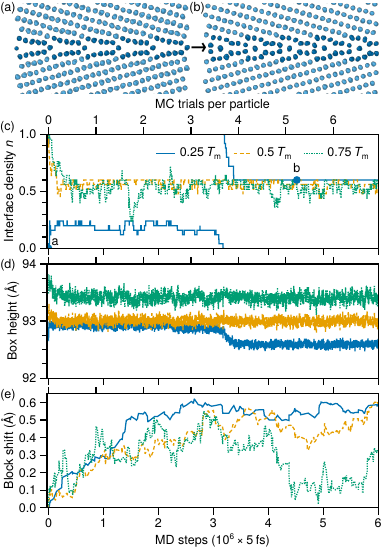}
\caption{
  \label{fig:gb1}
  $\mu PT$ MC equilibration of a $\Sigma27(552)[1\bar{1}0]$ tilt GB in W at three temperatures.
  (a) The initial structure, a $5\times 5$ tiling of the periodic unit, is a bicrystal with integer $(552)$ interface planes, i.e. $n = 0$.
  (b) The equilibrated structure at $T=0.25T_{\rm{m}}$, which is slightly lower energy than previous optimizations.
  (c) Evolution of $n$ vs the number of MC trials or MD steps.
  (d) Similarly, the cell length normal to the boundary, which defines volume, due to normal displacements of the upper bounding block.
  (e) Displacement of the bounding block parallel to the boundary.
}
\end{figure}

Indeed, $\mu PT$ MC is particularly valuable for simulating complex, extended defects such as crystal GBs.
Atomistic GBs are usually created by joining two misoriented crystals, but this construction does not generally correspond to physical equilibrium.
Crucially, the structure of a given boundary will change as atoms are added or removed from its core \cite{tasker83,phillpot92,vonalfthan06}.
GBs form distinct phases at different atomic densities, which are usually represented as fractions of the crystal plane parallel to the interface \cite{frolov13,frolov13a}.
The symmetric tilt and twist GBs considered in this study have only one such plane, which contains $N_{\parallel}$ atoms in the area of a simulation.
Since equilibrium GB structure is unaffected by the addition or removal of a complete parallel plane, the atomic density is represented as
\begin{equation}
  n = (N \bmod  N_{\parallel} )/ N_{\parallel}.
\end{equation}
A general definition of $n$ that includes asymmetric boundaries is given elsewhere \cite{winter25}. 

Conventional $NPT$ simulations with periodic boundary conditions artificially constrain $n$, preventing most GB phase transformations.
This commonly employed approach often creates unphysical GBs such as amorphous phases at unrealistically low temperatures \cite{vonalfthan06,frolov13}.
At high temperatures, open surface boundary conditions can allow atoms to diffuse in and out of a GB \cite{frolov13,frolov13a}, but this more qualitative approach requires large areas to mitigate surface effects and long simulation times that are limited by the rate of diffusion.
Rigorous studies of GB structure have instead focused on exploring 0\,K ground states with different numbers of atoms \cite{zhu18,banadaki18,chen24}.
These methods have proven successful for compositionally simple systems but cannot account for vibrational or configurational entropy, which affect interfaces as much as bulk phases.
$\mu PT$ MC can overcome all of these limitations by exchanging GB atoms with a reservoir representing an equilibrium bulk crystal.
As discussed above, the crystal's chemical potential can be accurately approximated by integrating $g_0=G_0/N_0$ from a reference Einstein crystal \cite{frenkel84,freitas16}.

Since GB volume is not known \textit{a priori}, the dimension perpendicular to the interface must also vary according to the pressure of the crystal.
However, the periodic dimensions along the boundary should be fixed at the bulk lattice parameter to prevent contraction from interface stress.
As in previous studies \cite{freitas18,chen24}, the GB region was bounded with rigid crystalline blocks, which were displaced as distinct MC trials (see Appendix \ref{app:muPT}).
The remainder of the study demonstrates the capabilities of $\mu PT$ MC by application to W and Cu GBs with complex atomic structures that have been carefully studied by previous investigations.
These GBs cannot be predicted with conventional techniques as they require density optimization, exhibit large-area reconstructions, and feature intricate atomic arrangements that deviate significantly from simple constructions based on coincidence site lattices.

The $\Sigma27(552)[1\bar{1}0]$ W GB provides a particularly intriguing example: for the interatomic potential of Ref. \cite{marinica13}, a genetic algorithm predicted many distinct but nearly degenerate ground states corresponding to a $2\times3$ tiling of the periodic unit with $n = 0.5$, which could reflect aperiodicity \cite{frolov18}.
Figure \ref{fig:gb1} reports $\mu P T$ simulations of this GB, starting from the $n=0$ construction depicted in (a), which is a $5\times5$ tiling of the periodic unit (see \cite{NOTE1} for details).
Simulations at a quarter, half, and three-quarters of the experimental melting temperature ($T_{\rm{m}}=3695$\,K) all quickly converged to the equilibrium $n=0.6$ structure shown in Fig. \ref{fig:gb1}(b) via the paths drawn in \ref{fig:gb1}(c).
Equilibration of the bounding blocks simultaneously led to reductions in volume, which are plotted in Fig. \ref{fig:gb1}(d), and consistent translations along the interface, which are shown in \ref{fig:gb1}(e).
The $n=0.6$ structure appears very similar to the previously identified $n=0.5$ ground states, but it has about half a percent lower surface energy when relaxed at 0\,K.
The improvement is likely attributable to the ability of MC simulations to efficiently explore a larger, more flexible simulation cell.

\begin{figure}
  \includegraphics[width=\linewidth]{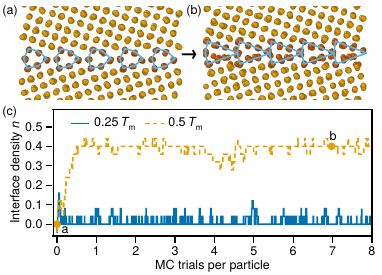}
\caption{
  \label{fig:gb2}
  $\mu PT$ MC equilibration of a $\Sigma 5(310)[001]$ tilt GB in Cu.
  (a) The initial ``normal kite" structure, a $5\times 5$ tiling of the periodic unit, is the 0\,K ground state with $n =0$.
  (b) The ``split-kite" structure is vibrationally stabilized at 0.5\,$T_{\rm{m}}$ with $n = 0.4$, in agreement with free energy calculations \cite{freitas18}.
  (c) Evolution of $n$ during simulations, showing a phase transformation at 0.5\,$T_{\rm{m}}$ and the metastability of the original phase at 0.25\,$T_{\rm{m}}$.
}
\end{figure}

$\mu PT$ MC can also equilibrate GBs with finite-temperature phases that differ from the 0\,K ground state.
According to the interatomic potential of Ref. \cite{mishin01}, a $\Sigma5(310)[001]$ Cu tilt boundary exhibits a first-order phase transition from an $n=0$ kite phase to an $n=0.4$ split-kite phase at 180\,K due to vibrational stabilization.
This transformation was previously established only by comparing free energy calculations for multiple enumerated structures \cite{freitas18}.
Figure \ref{fig:gb2} shows how $\mu PT$ MC simulations at 0.5\,$T_{\rm{m}}$ ($T_m = 1358$\,K) transform $n=0$ kites, which are depicted in Fig. (a), to the expected $n=0.4$ split kites shown in Fig. \ref{fig:gb2}(b), successfully capturing the phase transition.
However, the original $n=0$ kite structure remains stable at 0.25\,$T_{\rm{m}}$, albeit with notable fluctuations toward $n=0.4$ that indicate metastability.
While this behavior may reflect physical kinetics, $\mu PT$ MC, like many other methods, can struggle to escape local minima.

The direct equilibration of finite-temperature GB phases that were previously identified only by enumerating 0\,K structures represents a significant advance in interface structure prediction.
This capability relies on two key techniques.
First, the introduction of fictitious particles sets up a biasing scheme that greatly improves sampling efficiency over random insertions and deletions, particularly in heterogeneous structures like interfaces.
Gradually transforming these particles through Hamiltonian trajectories then enables moves, especially insertions, that would otherwise be impossible.
The method can be extended to simulate interfaces in multicomponent systems, which require finite-temperature sampling to describe chemical disorder and are generally too complex for exhaustive explorations of ground-state structures.
Basic configurational sampling could be incorporated by adding trials that swap the type of a real atom as in a semigrand canonical ensemble.

\begin{acknowledgments}
We thank T. Oppelstrup for helpful discussions.
This work was performed under the auspices of the U.S. Department of Energy by Lawrence Livermore National Laboratory (LLNL) under Contract No. DE-AC52-07NA27344.
The project was supported by the U.S. Department of Energy, Office of Science under an Office of Fusion Energy Science Early Career Award and partly supported by the LLNL Laboratory Directed Research and Development program under project tracking code 22-SI-007.
Computational resources were provided by the LLNL Institutional Computing Grand Challenge program and the Oak Ridge Leadership Computing Facility at the Oak Ridge National Laboratory, which is supported by the U.S. Department of Energy, Office of Science under Contract No. DE-AC05-00OR22725.
\end{acknowledgments}

\appendix

\section{$\mu PT$ calculations}
\label{app:muPT}
The $\mu PT$ partition function factors the implicit $V^N$ in $\bm{x}^N$, leaving dimensionless scaled coordinates, $\bm{s}^N$:
\begin{equation}
  Z_{\mu PT} = \sum_{N=0}^M \frac{1}{N!} \frac{1}{\Lambda^{3N}} \iint e^{-\beta \left(U(\bm{s}^{N\subset M},V) + PV - \mu N\right)} V^N d\bm{s}^N dV.
\end{equation}
Therefore, $MPT$ systems were sampled with
\begin{equation}
  \pi(\bm{s}^{N \subset M},V) = \frac{(M-N)!}{M!} \frac{V^{N-M}}{\Lambda^{3N}} e^{-\beta\left(U(\bm{s}^{N \subset M},V) +PV - \mu N\right)}.
\end{equation}
For simplicity, volume was sampled independently from $N$ and $\bm{s}^M$ using
\begin{equation}
  \frac{\pi(\bm{s}^{N \subset M},V_b)}{\pi(\bm{s}^{N \subset M},V_a)} = \left(\frac{V_b}{V_a} \right)^N e^{-\beta\left( U(\bm{s}^{N\subset M},V_b) - U(\bm{s}^{N\subset M},V_a) + P(V_b-V_a)\right)}
\end{equation}
and uniform trial probability densities, $W_{ba}^{V}/W_{ab}^{V} = 1$.
Equations for volume-conserving trials are unchanged.

In $\mu PT$ GB simulations, the directions normal to the GB were bounded with rigid crystalline regions that defined the thermodynamic volume and elastically reflected encroaching particles.
Depending on its position, $\bm{r}_b$, a bounding block contributes an additional potential energy term of $U_b(\bm{r}_b,\bm{x}^{N \subset M})$, which is treated as fixed $U_b(\bm{x}^{N \subset M})$ in Hamiltonian trajectories.
Volume moves were attempted by randomly displacing the upper block along the boundary normal within a range of fixed probability while scaling the coordinates of particles between the blocks.
In order to account for variations in relative grain translations, additional MC trials displaced the upper block parallel to the interface plane within similar fixed intervals.

\section{Point defect free energy model}
\label{app:ideal}
The $\mu PT$ simulations of point defects, which are exact in principle, can be further analyzed by comparison to a simple theoretical model.
A perfect crystal with $N_0$ sites has ideal free energy $g_0 N_0$.
The addition of $N_v = N_0 - N$ noninteracting vacancies contributes an ideal configurational entropy of $k_B \log {N_0 \choose N_v}$.
For a small number of vacancies under zero pressure, the change in free energy per vacancy can be approximated as $g_v - g_0$ for some $g_v$.
With these simplifications, 
\begin{equation}
  \label{eq:F}
   G = N g_0 + N_v g_v - k_B T \biggl[ N \log \biggl(\frac{N_v}{N} \biggr) - N_0 \log \biggl( \frac{N_v}{N_0} \biggr) \biggr].
\end{equation}
Taking $\mu = (\partial G / \partial N)_{T,P}$ and rearranging terms provides the vacancy concentration,
\begin{equation}
    \label{eq:N}
  \frac{N}{N_0} = \frac{1}{1+e^{-(\mu - g_0 + g_v)/k_B T }}.
\end{equation}
Given $g_0$, a value of $g_v$ can be extracted from a single calculation of $N(\mu)$; from $\mu=-4.07$\,eV, $g_v = 1.053$\,eV.
While interstitials can occupy additional sites, the same exercise implies an interstitial formation free energy of $g_i=1.890$\,eV at $\mu=-3.5$\,eV.
Using these values, Figs. \ref{fig:solid}(a) and \ref{fig:solid}(b) plot $\langle N \rangle$ and $g$ according to the model given by Eqs. (\ref{eq:N}) and (\ref{eq:F}).

\end{document}